\documentclass[pra, twocolumn, floatfix, superscriptaddress]{revtex4} 
\usepackage{graphicx}
\usepackage{color}
\usepackage{amsmath, amsfonts, amssymb, bm}
\begin{document}
\title{Relative-phase dependence of dynamically assisted electron-positron pair creation \\
in the superposition of strong oscillating electric-field pulses}

\author{J.~Bra{\ss}}
\address{Institut f\"ur Theoretische Physik I, Heinrich-Heine-Universit\"at D\"usseldorf, Universit\"atsstra{\ss}e 1, 40225 D\"usseldorf, Germany}
\author{D.~M.~M\"uller}
\address{Institut f\"ur Theoretische Physik I, Heinrich-Heine-Universit\"at D\"usseldorf, Universit\"atsstra{\ss}e 1, 40225 D\"usseldorf, Germany}
\author{S.~Villalba-Ch\'avez}
\address{Institut f\"ur Theoretische Physik I, Heinrich-Heine-Universit\"at D\"usseldorf, Universit\"atsstra{\ss}e 1, 40225 D\"usseldorf, Germany}
\author{K.~Krajewska}
\address{Institute of Theoretical Physics, Faculty of Physics, University of Warsaw, 
Pasteura 5, 02-093 Warsaw, Poland}
\author{C.~M\"uller}
\address{Institut f\"ur Theoretische Physik I, Heinrich-Heine-Universit\"at D\"usseldorf, Universit\"atsstra{\ss}e 1, 40225 D\"usseldorf, Germany}
\date{\today}
\begin{abstract}
Production of electron-positron pairs in the superposition of oscillating electric-field pulses with largely different frequencies is studied, focussing on the impact of relative phases between the pulses. Various field configurations are considered: superpositions of either two or three pulses of equal duration as well as combinations of a long low-frequency and a short high-frequency pulse. We show that the relative phase of superimposed high-frequency modes can exert a sizeable effect on the total numbers of produced pairs, enhancing them by about 10--30\% for the considered field parameters.
\end{abstract}

\maketitle

\section{Introduction}
In the presence of certain classes of electromagnetic fields the quantum vacuum 
can become unstable, leading to the production of electron-positron pairs
\cite{Review1, Review2, Review3, Review4, Review5}. The interest in this fundamentel 
phenomenon has strongly been revived in recent years, because dedicated experiments 
on field-induced pair production are under way at several high-intensity 
laser laboratories worldwide, including the Extreme-Light Infrastructure 
\cite{ELI}, Rutherford Appleton Laboratory \cite{RAL}, European X-Ray Free-Electron 
Laser \cite{LUXE}, Stanford Linear Accelerator Center (SLAC) \cite{E320} or
Center for Relativistic Laser Science \cite{CoReLS}. 

Oscillating electric fields are often applied in theoretical considerations 
as simplified field configurations to model laser pulses. This is because a 
standing laser wave, generated by two counterpropagating laser pulses, 
approaches to an oscillating, purely electric field in the vicinity 
of the wave's electric field maxima. Provided that the characteristic pair 
formation length is much smaller than the laser wavelength and focusing scale,
this kind of approximation is therefore applicable. 

In the seminal studies \cite{Brezin,Popov,Mostepanenko,Gitman} of pair production 
in monofrequent electric fields with infinite duration, various qualitatively 
different interaction regimes were found. They are divided by the dimensionless 
parameter $\xi = |e| E_0 /(mc\omega)$, where $E_0$ is the field amplitude and 
$\omega$ the field frequency; further we use the electron charge $e<0$, 
electron mass $m$, and speed of light $c$. The production probability for 
$\xi \ll 1$,  is given by a perturbative power law in the field amplitude $E_0$. 
Instead, for $\xi \gg 1$ with $E_0\ll E_{\rm cr}$, it possesses a nonperturbative 
exponential dependence on the ratio $E_{\rm cr}/E_0$, which closely resembles 
the Schwinger effect in a constant electric field \cite{Review1, Review2, Review3, 
Review4}. Here, $E_{\rm cr}=m^2c^3/(|e|\hbar)\approx 1.3\times 10^{16}$\,V/cm 
denotes the critical field strength. The nonperturbative regime of intermediate 
coupling strengths $\xi \sim 1$ lies in between these asymptotic behaviors, 
where analytical treatments of the problem are particularly difficult.

Beyond that, by considering in more recent years electric field pulses of finite 
duration \cite{AdP2004,subcycle,Mocken,GrobeJOSA} and manifold shapes 
\cite{Kohlfurst2013, Plunien2017, Grobe2019, Blinne, Bauke, Francois2017, Xie2017, 
Kohlfurst2019, Dunne2012, modulation, grating, Granz}, including frequency chirps 
\cite{Dumlu2010, Alkofer2019, Grobe2020} and spatial inhomogeneties \cite{Ruf, 
Alkofer, Dresden, Schutzhold-inhom, Kohlfurst2020, Kohlfurst2022}, a very rich 
phenomenology of the pair production was found, with the process rate depending 
sensitively on the precise form of the field. Field structures that are able to 
enhance the number of created pairs, are of special interest because they can help 
to facilitate an experimental observation of the process. 

Enhancement effects can arise, in particular, when two (or more) oscillating fields 
of different frequency are superimposed. On the one hand, when a weak high-frequency 
field mode is superposed on a strong low-frequency field background, a large
enhancement of pair production by several orders of magnitude can occur 
through the dynamically assisted Schwinger effect \cite{Schutzhold2008, Orthaber, 
Grobe2012, Akal, Opt, slit, Kampfer-survey, Kampfer-EPJD, Torgrimsson, Kampfer-EPJA, 
Selym-PRD, Folkerts}. This effect has also been studied in trifrequent fields 
\cite{doubly-assisted, Kampfer-EPJA} and for the nonlinear Bethe-Heitler \cite{DiPiazza, 
Augustin-PLB} and Breit-Wheeler \cite{Jansen-PRA} processes. On the other 
hand, coherent enhancements due to two-pathway quantum interferences arise when the 
field frequencies are commensurate, as was demonstrated for nonlinear Bethe-Heitler 
\cite{Krajewska, Augustin-PRA, Roshchupkin} and Breit-Wheeler \cite{Fofanov, Jansen-Proc} 
pair production. In this scenario, also a characteristic dependence on the relative 
phase between the field modes occurs \cite{Fofanov,Krajewska,Brass2020}. We note that 
a phase dependence of pair production was recently also revealed in the field of an 
atomic nucleus and a standing wave \cite{Grobe-phase}.

In the present paper, we study dynamically assisted electron-positron pair creation
in oscillating electric fields, where one (or two) high-frequency pulses of 
relatively low amplitude ($\xi\ll 1$, $\hbar\omega\sim mc^2$) are combined with 
a strong pulse of rather low frequency ($\xi\approx 1$, $\hbar\omega\ll mc^2$). 
Our main goal is to analyze the dependence of the {\it total} number of produced pairs 
on the relative phase(s) between the field pulses. This way, we extend previous 
results \cite{Akal, Kampfer-EPJA}, where a fixed value of the relative phase was 
assumed, and a recent study \cite{Folkerts}, where momentum distributions but no 
total pair numbers were obtained. The guiding question shall be as to which extent 
the mechanisms of dynamical assistence and coherent multiple-pathway interference 
can be combined with each other to further increase the resulting pair yields. 

Our paper is organized as follows. In Sec.\,II we briefly present three different 
computational schemes in the form of coupled differential equations, that 
have been established in the literature to calculate the momentum-dependent 
probabilities for pair creation in time-varying electric field pulses.
By way of example we show that the predictions of all three schemes agree 
with each other, which -- to our knowledge -- has not been shown before. 
In Sec.\,III we analyze the relative-phase dependence of dynamically assisted
pair production in various field configurations, comprising bifrequent (Sec.\,III A)
and trifrequent (Sec.\,III B) setups of uniform pulse duration as well as 
superpositions of an ultrashort assisting pulse onto a strong main pulse (Sec.\,III C). 
Conclusions are given in Sec.\,IV. Relativistic units with $\hbar = c = 1$ are used
in the following.

\section{Computational framework}
In our treatment of pair production in the superposition of oscillating electric field pulses, the fields are chosen to be linearly polarized in $y$-direction. Applying the temporal gauge, the total field $\vec E(t)=-\dot{\vec A}(t)$ can be described by a vector potential $\vec{A}(t) = A(t)\vec{e}_y$ of the form
\begin{eqnarray}
A(t) = \sum_{j = 1}^{K} A_j(t)\,F_j(t)
\label{A}
\end{eqnarray}
\color{black}
with
\begin{eqnarray}
A_j(t) = \frac{m\xi_j}{e}\,\sin(\omega_j t + \varphi_j)\ .
\end{eqnarray}
The envelope functions $F_j(t)$ have compact support on $[0,T_j]$, with sin$^2$-shaped turn-on and turn-off increments lasting $\delta_j$ cycles each and a constant plateau of unit height in between. Here, $T_j=\frac{2\pi}{\omega_j}(N_j+2\delta_j)$ denote the pulse durations and $N_j$ the number of plateau cycles. The electric field amplitudes of the modes are $E_j = m\xi_j\omega_j/|e|$ and $\varphi_j$ denote the relative phases. The number of superposed field pulses shall be either $K=2$ (in Secs.\,III A and C) or $K=3$ (in Sec.\,III B).

Various approaches have been used in the literature to derive coupled systems of ordinary differential equations (ODEs) from which the probability for pair production in an oscillating electric field can be obtained \cite{Mostepanenko, Gitman, Mocken, Akal, Kampfer-EPJD, Kampfer-EPJA, Hamlet, non-Markovian, Schmidt-1999}. We will briefly describe three of them and compare their predictions. Since Noether's theorem implies momentum conservation in spatially homogeneous fields, it is possible to treat the subspace associated with a certain momentum vector $\vec p$ separately. The rotational symmetry of the problem about the field axis allows us to parametrize the momentum as $\vec p = (p_x,p_y,0)$ with a transversal component $p_x$ and a longitudinal component $p_y$. In addition, there exists a conserved spin-like operator \cite{Mocken, Hamlet} that permits the separation of the problem into two spin semi-spaces.

\subsection{Dirac equation approach}
Starting from the time-dependent Dirac equation, the following representation was derived in Mocken {\it et al.}~\cite{Mocken} (see also \cite{Hamlet}):
\begin{eqnarray}
\dot{f}_{_{\rm M}}(t) &=& \kappa(t)f_{_{\rm M}}(t) + \nu(t)g_{_{\rm M}}(t)\ ,\nonumber\\
\dot{g}_{_{\rm M}}(t) &=& -\nu^*(t)f_{_{\rm M}}(t) + \kappa^*(t)g_{_{\rm M}}(t)\ ,
\label{system1}
\end{eqnarray}
with 
\begin{eqnarray}
\kappa(t) &=& ieA(t)\,\frac{p_y}{p_0}\ ,\nonumber\\
\nu(t) &=& -ieA(t)\,e^{2ip_0 t}\,\left[ \frac{(p_x-ip_y)p_y}{p_0(p_0+m)} + i\, \right] .
\label{nu}
\end{eqnarray}
It is obtained when an ansatz of the form $\psi_{\vec p}(\vec r,t) = f_{_{\rm M}}(t)\, \phi_{\vec p}^{(+)}(\vec r,t) + g_{_{\rm M}}(t)\, \phi_{\vec p}^{(-)}(\vec r,t)$ is inserted in the Dirac equation. Here, $\phi_{\vec p}^{(\pm)}\sim e^{i(\vec p\cdot\vec r \mp p_0 t)}$, with $p_0=\sqrt{{\vec p}^{\,2}+m^2}$, denote free Dirac states with momentum $\vec p$ and positive or negative energy. The time-dependent coefficients $f_{_{\rm M}}(t)$ and $g_{_{\rm M}}(t)$ describe the occupation amplitudes of a positive-energy and negative-energy state, accordingly. The system \eqref{system1} is solved with the initial conditions $f_{_{\rm M}}(0)=0$, $g_{_{\rm M}}(0)=1$. At time $T$ when the fields have been switched off, $f_{_{\rm M}}(T)$ represents the occupation amplitude of an electron state with momentum $\vec p$, positive energy $p_0$ and certain spin projection. Taking the two possible spin degrees of freedom into account, we obtain the probability for creation of an electron with energy $p_0$ and momentum components $p_x, p_y$ as 
\begin{eqnarray}
W(p_x,p_y) = 2\,|f_{_{\rm M}}(T)|^2\ .
\label{W}
\end{eqnarray}
The associated positron has the momenta $-p_x$ and $-p_y$.

\subsection{Quantum kinetic approaches}
Alternative ODE systems to describe the pair creation process in an electric field were derived from the corresponding quantum kinetic Boltzmann-Vlasov equation by applying the Bogolyubov transformation method. One of them can be found in Akal {\it et al.}~\cite{Akal} and has the following form:
\begin{eqnarray}
i\dot{f}_{_{\rm A}}(t) &=& a(t)f_{_{\rm A}}(t) + b(t)g_{_{\rm A}}(t)\ ,\nonumber\\
i\dot{g}_{_{\rm A}}(t) &=& b^*(t)f_{_{\rm A}}(t) - a(t)g_{_{\rm A}}(t)\ ,
\label{system2}
\end{eqnarray}
with 
\begin{eqnarray}
a(t) &=& w(t) + \frac{eE(t)p_x}{2w(t)[w(t)+m]}\ ,\nonumber\\
b(t) &=& \frac{1}{2}\frac{eE(t)\epsilon_\perp}{w(t)^2}\exp\left[-i\arctan \left(\frac{p_xq_\parallel}{\epsilon_\perp^2+w(t)m}\right)\right]\,,
\label{a}
\end{eqnarray}
where $w(t) = \sqrt{\epsilon_\perp^2 + q_\parallel^2}$ is the instantaneous particle energy, $\epsilon_\perp = \sqrt{m^2 + p_x^2}$ its transverse part and $q_\parallel = p_y - eA(t)$ the longitudinal kinetic momentum. From the coupled ODE system \eqref{system2}, the probability for creation of an electron with definite momentum is obtained analogously to Eq.~\eqref{W} as $2\,|f_{_{\rm A}}(T)|^2$.

Another possible system of equations was used in Otto {\it et al.}~\cite{Kampfer-EPJA}; it has the following form:
\begin{eqnarray}
\dot{f}_{_{\rm O}}(t) &=& Q(t)u_{_{\rm O}}(t)\ ,\nonumber\\
\dot{u}_{_{\rm O}}(t) &=& Q(t)[1 - f_{_{\rm O}}(t)] - 2w(t)v_{_{\rm O}}(t)\ ,\nonumber\\
\dot{v}_{_{\rm O}}(t) &=& 2w(t)u_{_{\rm O}}(t)\ ,
\label{system3}
\end{eqnarray}
with the same $w(t)$ as above,  
\begin{eqnarray}
Q(t) = \frac{eE(t)\epsilon_\perp}{w(t)^2}\,,
\label{Q}
\end{eqnarray}
and the initial conditions $f_{_{\rm O}}(0)=u_{_{\rm O}}(0)=v_{_{\rm O}}(0)=0$.
Here, $f_{_{\rm O}}(t)$ represents the full one-particle distribution function, which includes a sum over the spins. At $t=T$, when the field is turned off, it represents the probability for production of an electron with momenta $p_x$ and $p_y$.

\subsection{Computation of total number of electron-positron pairs}
From the momentum-dependent production probability $W(p_x,p_y)$ one can calculate the total number of produced pairs by integration over the momentum space. Due to the cylindrical symmetry of the problem, the integral to be evaluated numerically can be written as \cite{Mocken, Akal}
\begin{eqnarray}
\mathcal{N} \approx \frac{1}{4\pi^2} \int_{-p_y^{\rm (max)}}^{p_y^{\rm (max)}}\!\int_{0}^{p_x^{\rm (max)}}\!\!\! W(p_x,p_y)\,p_x\,dp_x dp_y
\label{N}
\end{eqnarray}
where $p_x^{\rm (max)}$ and $p_y^{\rm (max)}$ denote sufficiently large boundaries to cover the region where the production is sizeable. These boundaries have to be adjusted to the chosen field parameters. In our calculations, $p_x^{\rm (max)}=p_y^{\rm (max)}=3m$ will be taken throughout.
Equation~\eqref{N} gives the number of pairs produced in a Compton volume $(1/m)^3$.

The following numerical results have been simulated with the help of a custom C++ code which has been benchmarked with the help of a standard ODE solver in order to insure the accuracy of the results. The custom code is using a time step of $\Delta t = 0.01m^{-1}$. The momentum resolution of the probability planes is chosen as $\Delta p_{x,y} = 0.01m$. The phase step size will be $\Delta \varphi_i = 2\pi/72$ and the time delay stepsize in Sec.III.C is $\delta t = 0.25m^{-1}$.

The field frequencies entering the pair production setups considered in this paper will be chosen to lie in the range $\omega\sim (0.1$ - $1)m$. While for the weak field modes, such high frequencies are physically motivated by the underlying mechanism of dynamical assistance \cite{Schutzhold2008, Orthaber, Grobe2012, Akal, Opt, slit, Kampfer-survey, Kampfer-EPJD, Torgrimsson, Kampfer-EPJA, Selym-PRD, Folkerts}, for the strong mode a rather large frequency is chosen for reasons of computational feasibility. We note that, in this situation, a purely time-dependent field represents a simplified model (rather than a close approximation) for the electromagnetic fields of a standing laser wave. Corresponding differences in the field-induced pair creation -- arising from the spatial dependence and magnetic component 
of laser fields -- have been revealed in Refs.~\cite{Ruf, Alkofer, Dresden, Schutzhold-inhom, Kohlfurst2020, Kohlfurst2022}.

\section{Numerical results}

\subsection{Bifrequent pulses of equal length}
The bifrequent oscillating electric-field (bOEF) pulses in this section have the general form 
\begin{eqnarray}
A_{\text{bOEF}}(t) = \left[A_1(t,\varphi_1) + A_2(t,\varphi_2)\right] F_1(t)\,,
\end{eqnarray}
with a common envelope function $F_1(t)$. Consequently, both field modes have the same temporal duration. 

\begin{figure}[b]
\centering
\includegraphics[width=\linewidth]{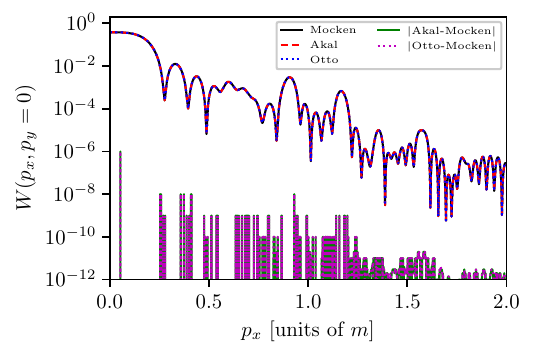}

\vspace{-0.25cm}
\caption{Transversal momentum distributions of electrons produced with $p_y=0$ in a bOEF pulse with parameters $\xi_1 = 1.0$, $\xi_2 = 0.1$, $\omega_1 = 0.3m$, $\omega_2 = 1.24385m$, $N_1 = 4$, and $\varphi_1=\varphi_2=0$. The black solid curve shows the result of Eq.~\eqref{system1}, the red dashed curve relies on Eq.~\eqref{system2} and the blue dotted curve displays the outcome of Eq.~\eqref{system3}. The green solid curve represents the absolute difference between the results of Eqs.~\eqref{system2} and \eqref{system1}, whereas the magenta dotted curve shows the absolute difference between the results of Eqs.~\eqref{system3} and \eqref{system1}.}
\label{fig:Compare}
\end{figure}

In Fig.~\ref{fig:Compare} we compare the predictions from the three ODE systems given in Secs.\,II A and B for pair production in a bOEF pulse. The field amplitudes and frequencies have been chosen in line with those considered in Ref.\,\cite{Akal}. We note that the field-dressed mass of the particles in this configuration amounts to $m_\ast\approx 1.21m$. The highest production probabilities are achieved for $p_x\approx 0$, where the pairs are created predominantly by the absorption of four 'small' photons $\omega_1$ and one 'big' photon $\omega_2$ (see Table\,I in \cite{Akal}), leading to $4\omega_1+\omega_2\approx 2m_\ast$.

With a high degree of accuracy, all three ODE systems yield identical pair production probabilities. While this outcome is to be expected, to the best of our knowledge a corresponding consistency check has not yet been performed.
In the following, we use the ODE system of Eq.~\eqref{system2}, because it has a similar structure as Eq.~\eqref{system1} and follows from a quantum kinetic approach like Eq.~\eqref{system3}. In this sense, it lies in between the other two ODE systems.

\medskip

We now turn to the phase dependence of the total pair production yields in bOEF pulses. The field amplitudes and frequencies in our first example are motivated by Ref.~\cite{Akal} where a strong enhancement of the pair yield due to dynamical assistance by the second field mode was found, restricting the consideration to fixed phase values of $\varphi_1=\varphi_2=0$. It is an intriguing question whether the total number of produced pairs can be further enhanced by adjusting the relative phase $\varphi_2$ of the assisting high-frequency mode. 

\begin{figure}[t]
\centering
\includegraphics[width=0.9\linewidth]{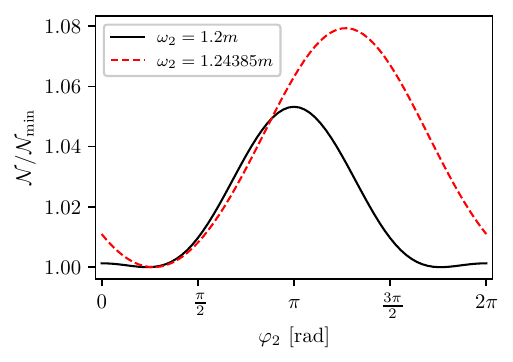}
\vspace{-0.25cm}
\caption{Total number of created pairs (per Compton volume), normalized to the minimum amount of produced pairs. Shown is the dependence on the relative phase $\varphi_2$ of a bOEF pulse with parameters $\xi_1 = 1.0$, $\xi_2 = 0.1$, $\omega_1 = 0.3m$ and $N_1 = 4$. The black solid curve corresponds to $\omega_2 = 1.2m$ and the red dashed curve to $\omega_2 = 1.24385m$. The minimal numbers of produced pairs are $\mathcal{N}_{\text{min}} = 1.12 \cdot 10^{-4}$ for $\omega_2 = 1.2m$ and $\mathcal{N}_{\text{min}} = 1.53 \cdot 10^{-4}$ for $\omega_2 = 1.24385m$, respectively.}
\label{fig:DynAssi}
\end{figure}

\begin{table}[b]
\caption{\label{tab:table1}
Total number $\mathcal{N}$ of electron-positron pairs for single electric-field modes and minimal number $\mathcal{N}_{\rm min}$ when the modes are combined to a bOEF pulse. The length of the pulses corresponds throughout to $N_1 = 4$ and $\delta = 0.5$.}
\begin{ruledtabular}
\begin{tabular}{lllll}
\textrm{$\xi_1$}&
\textrm{$\xi_2$}&
\textrm{$\omega_1 \left[m\right]$}&
\textrm{$\omega_2 \left[m\right]$}&
\textrm{$\mathcal{N}_{\rm (min)}$}\\
\colrule
1.0 & -   & 0.3 & -       & $2.25 \cdot 10^{-6}$ \\
-   & 0.1 & -   & 1.2     & $3.24 \cdot 10^{-5}$ \\
-   & 0.1 & -   & 1.24385 & $3.90 \cdot 10^{-5}$ \\ \hline
1.0 & 0.1 & 0.3 & 1.2     & $1.12 \cdot 10^{-4}$ \\
1.0 & 0.1 & 0.3 & 1.24385 & $1.53 \cdot 10^{-4}$ \\
\end{tabular}
\end{ruledtabular}
\end{table}

Figure~\ref{fig:DynAssi} shows our corresponding results, with the phase of the strong low-frequency mode set to zero. Note that here and henceforth, the curves have been normalized by dividing through the minimum of $\mathcal{N}(\varphi_2)$ as a function of $\varphi_2$. This facilitates a comparison of the shape of the curves and highlights the phase dependence.
In the considered field configurations, the phase dependence is found to be rather weak, leading to relative variations of the total pair number by a few percent only. When $\omega_2$ is exactly four times larger than $\omega_1$, the resulting graph is symmetric about $\varphi_2=\pi$ where the pair number is maximized. Otherwise, when the frequencies are practically incommensurate, a non-symmetric phase dependence arises, with the maximum shifted to $\varphi_2\approx 1.25\pi$.

As the numbers of produced pairs in Tab.~I show the presence of the assisting high-frequency mode strongly enhances the pair yield. In the combined field, the pair number is signficantly larger than the sum of the pairs produced in each mode separately.

\subsection{Trifrequent pulses of equal length}

Now we move to pair production in trifrequent oscillating electric-field (tOEF) pulses of the form
\begin{eqnarray}
A_{\text{tOEF}}(t) = \left[A_1(t,\varphi_1) + A_2(t,\varphi_2) + A_3(t,\varphi_3)\right] F_1(t).
\end{eqnarray}

\begin{figure}[t]
\centering
\includegraphics[width=0.9\linewidth]{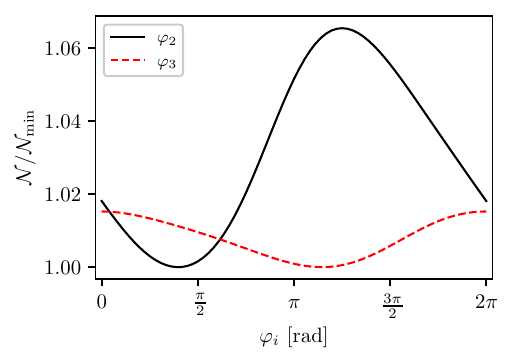}
\includegraphics[width=0.9\linewidth]{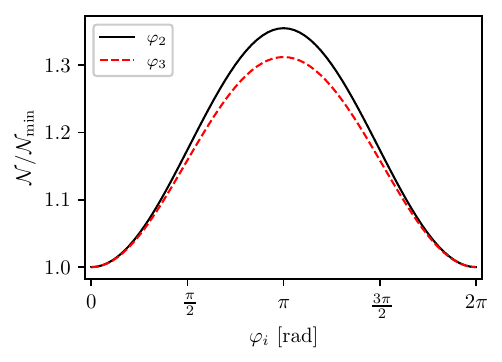}
\vspace{-0.25cm}
\caption{Total number of electron-positron pairs created in a tOEF pulse with $\xi_1 = 1.0$, $\xi_2 = 0.1$, $\xi_3 = 0.0027$, $\omega_1 = 0.3m$, $N_1 = 4$ and $\delta_1=0.5$, normalized to the minimum amount of produced pairs. The black solid curves represent the $\varphi_2$ dependence and the red dashed curves the $\varphi_3$ dependence. Top panel: $\omega_2 = 1.24385m$, $\omega_3 = 2.4877m$; here  the minimal amount of pairs produced is $\mathcal{N}_{\text{min}}(\varphi_2) = 1.96 \cdot 10^{-4}$ and  $\mathcal{N}_{\text{min}}(\varphi_3) = 1.97 \cdot 10^{-4}$ respectively.
Bottom panel: $\omega_2 = 1.2m$ and $\omega_3 = 2.4m$ are integer multiples of $\omega_1$. The minimal pair number is $\mathcal{N}_{\text{min}}(\varphi_2) = \mathcal{N}_{\text{min}}(\varphi_3) = 1.26 \cdot 10^{-4}$.}
\label{fig:DATC04}
\end{figure}

First we consider tOEF scenarios that are based on the bOEF parameters of Fig.~\ref{fig:DynAssi}, to which a third field mode is added whose frequency is twice as large as $\omega_2$. As the top panel in Fig.~\ref{fig:DATC04} shows, the dependencies on the relative phases remain weak, when the strong-mode frequency $\omega_1$ is practically incommensurate to the higher field frequencies. The shape of the $\varphi_2$ dependence is very similar to the bOEF case of Fig.~\ref{fig:DynAssi}. The changes in the number of produced pairs are even weaker when the phase $\varphi_3$ is varied. Note that the phases $\varphi_i$, which are not varied, are set to zero.

The influence of the field phases substantially increases in the bottom panel of Fig.~\ref{fig:DATC04} where the field frequencies form integer ratios. Here the pair yield can be enhanced by more than 30\% when the field phase is chosen properly. For variation of both $\varphi_2$ and $\varphi_3$, the maximum appears at a phase value of $\pi$. The position of the maxima coincides with the commensurate bOEF case of Fig.~\ref{fig:DynAssi}, and also the shapes of the curves are very similar. 

For both field configurations in Fig.~\ref{fig:DATC04}, the enhancement of the pair yields due to the combination of the three field modes is substantial when compared with the effect of each single pulse (see Tab.~I and II). However, a comparison with the bOEF cases of Fig.~\ref{fig:DynAssi} reveals, that the impact of the third mode is rather moderate, as it leads to only a minor further enhancement.

\begin{table}[b]
\caption{\label{tab:table2}
Total number of electron-positron pairs for the third field mode of Fig.~\ref{fig:DATC04} alone (with $N_1 = 4$ and $\delta_1 = 0.5$).}
\begin{ruledtabular}
\begin{tabular}{lllll}
\textrm{$\xi_3$}&
\textrm{$\omega_3 \left[m\right]$}&
\textrm{$\mathcal{N}$}\\
\colrule
0.0027 &  2.4877 & $4.29 \cdot 10^{-5}$ \\
0.0027 &  2.4    & $3.78 \cdot 10^{-5}$ \\
\end{tabular}
\end{ruledtabular}
\end{table}

\medskip

Our second tOEF scenario relies on the field amplitudes and commensurate frequencies considered in \cite{Kampfer-EPJA}, where doubly assisted pair creation was studied. Since fixed field phases of $\varphi_1=\varphi_2=\varphi_3=0$ were taken there, the question arises again to which extent the number of produced pairs can be further enhanced by variation of the field phases. Especially the phases $\varphi_2$ and $\varphi_3$ of the assisting modes are of interest in this regard. 

\begin{figure}[tbh]
\centering
\includegraphics[width=0.82\linewidth]{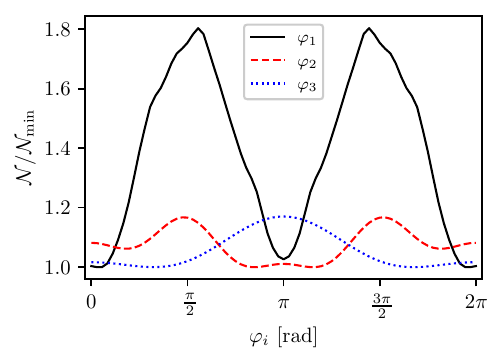}
\includegraphics[width=0.82\linewidth]{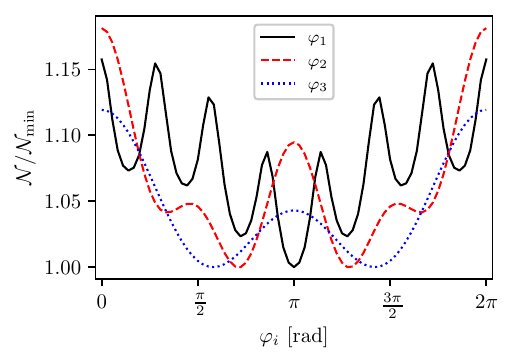}
\includegraphics[width=0.82\linewidth]{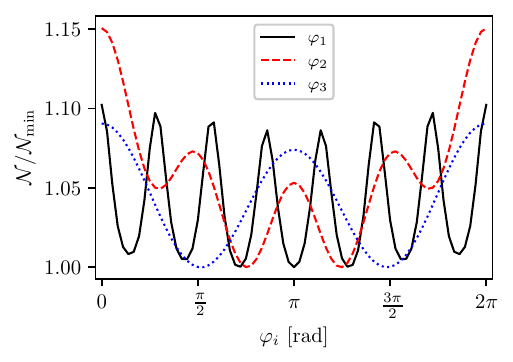}
\vspace{-0.25cm}
\caption{Total number of electron-positron pairs created in a tOEF pulse with $\xi_1 = 1.42$, $\xi_2 = 0.102$, $\xi_3 = 0.0102$, $\omega_1 = 0.07m$, $\omega_2 = 0.49m$, and $\omega_3 = 0.98m$,
normalized to the minimum amount of produced pairs. The black solid, red dashed and blue dotted curves show the dependencies on $\varphi_1$, $\varphi_2$ and $\varphi_3$, respectively (with the other phases, which are not being varied, set to zero). Top panel: $N_1 = 4$, $\delta_1=0.5$. The minimal numbers of produced pairs are $\mathcal{N}_{\text{min}}(\varphi_1) = 1.53 \cdot 10^{-8}$, $\mathcal{N}_{\text{min}}(\varphi_2) = 1.42 \cdot 10^{-8}$ and $\mathcal{N}_{\text{min}}(\varphi_3) = 1.51 \cdot 10^{-8}$. Middle panel: $N_1 = 8$, $\delta_1 = 1$. Here $\mathcal{N}_{\text{min}}(\varphi_1) = 2.86 \cdot 10^{-8}$, $\mathcal{N}_{\text{min}}(\varphi_2) = 2.81 \cdot 10^{-8}$, $\mathcal{N}_{\text{min}}(\varphi_3) = 2.96 \cdot 10^{-8}$. Bottom panel: $N_1 = 20$, $\delta_1 = 2$. Here $\mathcal{N}_{\text{min}}(\varphi_1) = 7.18 \cdot 10^{-8}$, $\mathcal{N}_{\text{min}}(\varphi_2) = 6.88 \cdot 10^{-8}$, $\mathcal{N}_{\text{min}}(\varphi_3) = 7.25 \cdot 10^{-8}$.}
\label{fig:tOEF2}
\end{figure}

Figure~\ref{fig:tOEF2} shows the phase dependencies for very short pulses with $N_1=4$ (top panel) and longer pulses with $N_1=8$ (middle panel) and $N_1=20$ (bottom panel). The duration of the turn-on and turn-off segments has been adjusted, accordingly, so that the pulses have a similar form in all three cases. The red dashed and blue dotted curves in the top panel reveal that the pair yields can be enhanced by about 20\% when the assisting field phase $\varphi_2$ or $\varphi_3$ is suitably chosen. 

For completeness we also show the dependence on the strong-field phase $\varphi_1$, which is very pronounced here. Setting it to $\varphi_1\approx 0.55\pi$ or $1.45\pi$, the number of pairs is almost doubled. For these optimum phase values, the vector potential $A_1$ becomes approximately cos-shaped. Nonlinear strong-field processes induced by very short pulses are generally known to be highly sensitive to the field phase. However, this kind of enhancement is not our main interest here, since variation of $\varphi_1$ also changes the number of pairs produced by the pulse $A_1$ alone.

Based on the results of the top panel in Fig.~\ref{fig:tOEF2}, we have also computed the dependencies on $\varphi_2$ and $\varphi_3$, when the strong-field mode is exactly cos-shaped, setting $\varphi_1=\frac{\pi}{2}$ in Eq.~\eqref{A}. In this case, the number of pairs can also be enhanced by about 20\% when the phase $\varphi_2$ is properly chosen, whereas the pair yield is almost independent of the phase $\varphi_3$ (not shown). 

The phase dependencies become more complex for the longer field durations and, as a general trend, the enhancement effect reduces (for all three phase variations). In the middle panel of Fig.~\ref{fig:tOEF2}, where the number of strong-field plateau cycles is taken as $N_1=8$, with turn-on and turn-off segments of $\delta=1$, optimal values of $\varphi_2=0$ and $\varphi_3=0$ are obtained. They generate a relative enhancement over the minimal pair yields of 12--18\%. The variation of $\varphi_1$ leads to seven maxima; this number coincides with the frequency ratio $\omega_2/\omega_1$. A similar behavior is seen in the bottom panel for the pulse length $N_1=20$, with relative enhancements of 9--15\%.

In all three cases shown in Fig.~\ref{fig:tOEF2}, the number of produced pairs is on the order of $10^{-8}$ to $10^{-7}$ and thus much larger than the pair numbers created by each single pulse individually, which lie here in the range $\sim 10^{-12}$--$10^{-10}$.

\subsection{Superposed short pulse onto long pulse}
\begin{figure}[t]
\centering
\includegraphics[width=0.85\linewidth]{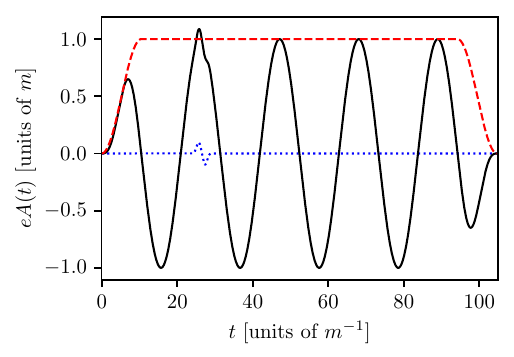}

\vspace{-0.15cm}
\caption{Total vector potential (black solid line), multiplied by $e$, of a weak ultrashort pulse superposed with some time delay on a strong background pulse, which is enclosed by a window function (red dashed line) with sin$^2$-shaped turn-on and turn-off segments. The blue dotted line shows $eA_2$ of the weak ultrashort pulse alone. The field parameters are $\xi_1 = 1.0$, $\xi_2 = 0.15$, $\omega_1 = 0.3m$, $\omega_2 = 1.24385m$, $N_1 = 4$, $N_2 = 0$, $\delta_1=\delta_2=0.5$ and $\Delta t = 24m^{-1}$.}
\label{fig:KPLP-scheme}
\end{figure}

So far, we have considered pair production in electric fields resulting from the superposition of two or three pulses of the same length. However, it is also possible to superpose pulses of different lengths. Then, similarly to a relative phase shift between pulses, the question arises to which extent a time shift between two superposed pulses influences the pair production process. Such a situation has recently been studied \cite{Folkerts} for an electric field with vector potential of the form
\begin{eqnarray}
A_{\rm LS}(t) = A_1(t)F_1(t) + A_2(t-\Delta t)F_2(t-\Delta t)\,.
\label{KPLP}
\end{eqnarray}
It describes the superposition of two pulses, each of which equipped with its own envelope function. The second pulse, which is assumed to be much shorter than the first pulse, contains a time delay $\Delta t$. The field configuration is illustrated in Fig.~\ref{fig:KPLP-scheme}. In order to guarantee that the shorter second pulse always overlaps in its entirety with the longer first pulse, we restrict the time delay to $\Delta t_{\rm max} = T_1 - T_2 = \frac{2\pi}{\omega_1}(N_1 + 1) - \frac{2\pi}{\omega_2}(N_2 + 1)$. 

\begin{figure}[b]
\centering
\includegraphics[width=0.9\linewidth]{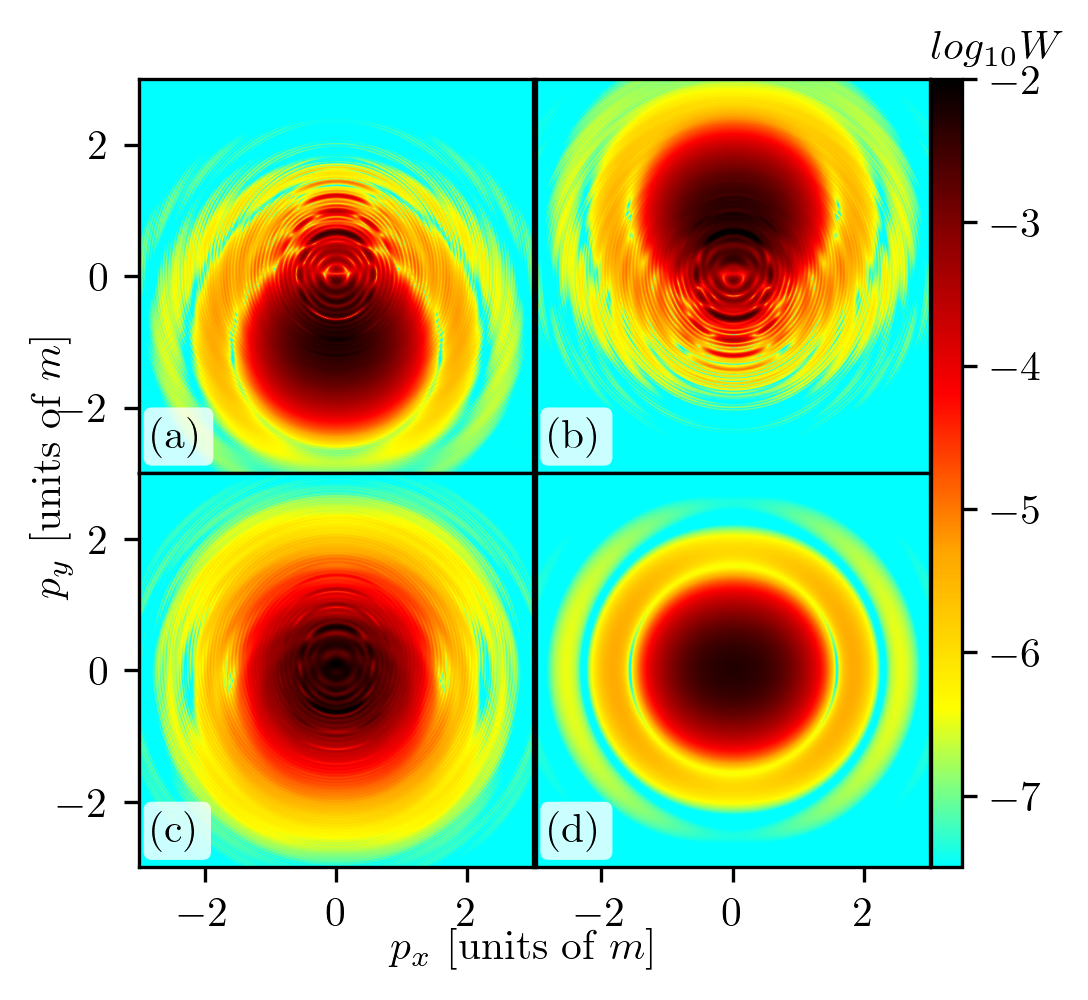}
\caption{Two-dimensional momentum distribution of electrons created in the two-pulse field configuration of Eq.~\eqref{KPLP} with parameters $\xi_1 = 1.0$, $\xi_2 = 0.15$, $\omega_1 = 0.3m$, $\omega_2 = 1.24385m$, $N_1 = 4$ and $N_2 = 0$. Panel (a) refers to a delay of $\Delta t = 14m^{-1}$, which leads to maximal displacement of the 'blob' in negative $p_y$-direction. Panel (b) shows the opposite case for $\Delta t = 22m^{-1}$. Panel (c) shows the result for an intermediate delay $\Delta t = 18m^{-1}$, where the 'blob' is located in the middle. For comparison, panel (d) shows the momentum distribution for an isolated short pulse (i.e., $A_1\equiv 0$ here).}
\label{fig:KPLP-2D}
\end{figure}

\begin{figure}[b]
\centering
\includegraphics[width=\linewidth]{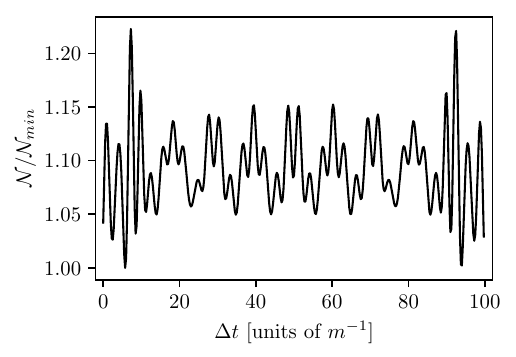}

\vspace{-0.15cm}
\caption{Total number of electron-positron pairs created in a two-pulse field configuration, normalized to the minimal pair yield given by $\mathcal{N}_{\text{min}} = 3.58 \cdot 10^{-5}$. The field parameters of the pulses are the same as in Fig.~\ref{fig:KPLP-2D} and the time delay is varied.}
\label{fig:KPLP-total}
\end{figure}

In the previous study \cite{Folkerts} the impact of the time delay on the one-dimensional momentum distributions of the created particles in longitudinal ($p_y$) and transversal ($p_x$) directions has been analysed. This analysis is extended here by investigating the full two-dimensional momentum distribution in Fig.~\ref{fig:KPLP-2D} and the total number of created pairs in Fig.~\ref{fig:KPLP-total}. Both figures display pronounced effects on the pair production when the time delay is varied.

Figure~\ref{fig:KPLP-2D} demonstrates that the momentum distribution of the created particles is very sensitive to the position of the short pulse $A_2$ on the background of pulse $A_1$. For the chosen parameters, the duration of the short pulse is $T_2\approx 5m^{-1}$; it solely consists of half a cycle for turn on and half a cycle for turn off, without plateau in between ($N_2=0$). When it is placed around a minimum of $eA_1(t)$, which is the case in panel (a), the distribution is strongly dominated by electrons with $p_y<0$. Instead, when the short pulse sits on a maximum of $eA_1(t)$ as in panel (b), predominantly electrons with $p_y>0$ are produced. By comparison with panel (d), which shows the momentum distribution when solely the short pulse $A_2$ is present, we realize that this 'blob' of electrons is moved downwards or upwards, depending on the applied pulse delay. When the short pulse surrounds a zero-crossing of $A_1(t)$, the blob lies in the middle, as panel (c) displays. 

In addition to the dominant blob in panels (a)-(c), a ring-like structure is visible in the center. It has the characteristic form of a momentum distribution arising in a monofrequent field (see, e.g., Fig.~11 in \cite{Mocken} or Fig.~1 in \cite{Brass2020}) and can be attributed to the background pulse $A_1(t)$. The position of this structure remains stationary when the time delay of the second pulse is varied. The rings are due to multiphoton resonances that arise when $n\omega_1$ (with an integer $n$) equals the energy of the created pair. The momentum distribution in panel (d) from the short pulse alone is much more continuous than the ring-like structure because of the extreme shortness of this pulse which corresponds to a large frequency bandwidth. In addition to the 'blob' generated by $A_2$ and the ring-like structure generated by $A_1$, the quantum interference between them leads to pronounced side fringes.

Figure~\ref{fig:KPLP-total} shows the (normalized) total number of produced pairs, as function of the applied time delay between the pulses. When the second pulse overlaps with either the turn-on or turn-off segment of the first pulse, the pair yield is particularly sensitive to the precise value of $\Delta t$. However, also when the ultrashort second pulse is superposed onto the plateau region of the first pulse, the time delay affects the pair production, leading to a relative variation of the yield by about 10\%. Somewhat similar regular structures of minima and maxima have been found in \cite{Folkerts} when the pair production probability for fixed momentum was considered (see Fig.~8 therein). However, for different values of momenta the sequence of minima and maxima looks quite differently. It is therefore interesting that, after integration over all momenta, similar regular structures are still visible in the total number of produced pairs.

The minimal amount of pairs produced in the two-pulse field configuration of Fig.~\ref{fig:KPLP-total} approximately coincides with the pair yield from the short pulse alone (see Tab.~III). At time delays where the number of produced pairs is enhanced by 15-20\% over the minimal amount, the pair yield from the superposed pulses is larger than the sum of the pair yields from each single pulse alone.

\begin{table}[htb]
\caption{\label{tab:table1}
Total number of electron-positron pairs for the single electric-field modes associated with Figs.~\ref{fig:KPLP-2D} and \ref{fig:KPLP-total}.}
\begin{ruledtabular}
\begin{tabular}{llllccll}
\textrm{$\xi_1$}&
\textrm{$\xi_2$}&
\textrm{$\omega_1 \left[m\right]$}&
\textrm{$\omega_2 \left[m\right]$}&
\textrm{$N_1$}&
\textrm{$N_2$}&
\textrm{$\delta$}&
\textrm{$\mathcal{N}$}\\
\colrule
1.0 & -    & 0.3 & -       & 4 & - & 0.5 & $2.25 \cdot 10^{-6}$ \\
-   & 0.05 & -   & 1.24385 & - & 0 & 0.5 & $3.57 \cdot 10^{-5}$ \\
\end{tabular}
\end{ruledtabular}
\end{table}

\section{Conclusion and Outlook}
The relative-phase dependence of dynamically assisted electron-positron pair production in various configurations of oscillating electric field pulses with largely different frequencies has been studied. To this end, superpositions of either two or three pulses of equal duration as well as combinations of a long low-frequency and a short high-frequency pulse were considered.

While it is known that the superposition of a weak high-frequency mode onto a strong low-frequency mode can enhance the pair yield enormously, we found that variation of the relative field phase can lead to additional enhancements of about 10--30\% for the considered field configurations and applied parameters. On the one hand, this represents an only moderate effect, which could have been expected since the frequencies required for dynamically assisted pair production are largely different and not necessarily commensurate (as opposed to the pronounced phase effects which can occur when a second harmonic is superposed onto a fundamental frequency). On the other hand, the additional enhancement effect appears sizeable in light of the fact that changing the relative phase of an assisting field mode can be quite easily accomplished. In particular, it does not require changing the total field energy contained in the assisting pulse.

In the present study, we performed one-dimensional analyses of the pair yields when one field phase is varied while the other phase(s) remain fixed. Beyond that, the effects of simultaneous variations of more than one field phase would be interesting to explore. This extended more-dimensional phase analysis, which would be computationally very demanding, is left for future studies.

\begin{acknowledgements}
This study was funded by the Deutsche Forschungsgemeinschaft (DFG) 
under Grant No.~392856280 within the Research Unit FOR 2783/2.
Computational support via the HPC system Hilbert was provided by 
the Zentrum f\"ur Informations- und Medientechnologie (ZIM) at 
Heinrich-Heine-Universit\"at D\"usseldorf.
\end{acknowledgements}


\end{document}